\documentclass[usenatbib]{mn2e}
\usepackage{graphicx}
\usepackage{abbreviations}
\footnotesize
\newdimen\digitwidth 
\setbox0=\hbox{\rm0}
\digitwidth=\wd0
\catcode`!=\active
\def!{\kern\digitwidth}
\normalsize

\title[Coronal Magnetic Field]
{Measurement of the electron density and magnetic field of the solar wind using millisecond pulsars}
\author[X. P. You et al.]
{X. P. You,$^1$\thanks{Email:yxp0910@swu.edu.cn}
W. A. Coles,$^2$
G. B. Hobbs,$^3$
R. N. Manchester$^3$
\\
$^1$ School of Physical Science and Techology, Southwest University,
Chongqing, 400715, China\\
$^2$ Electrical and Computer Engineering, University of California at San Diego, La Jolla, California, U.S.A \\
$^3$ CSIRO Astronomy and Space Science, Australia Telescope National Facility, P.O.~Box~76, Epping NSW~1710, Australia
}

\let\leqslant=\leq 
\date{}

\begin{document}

\maketitle
\newcommand{\setthebls}{
}
\setthebls
\begin{abstract}
  The magnetic field of the solar wind near the Sun is very difficult
  to measure directly.  Measurements of Faraday rotation of linearly
  polarized radio sources occulted by the solar wind provide a unique
  opportunity to estimate this magnetic field, and the technique has
  been widely used in the past.  However Faraday rotation is a path
  integral of the product of electron density and the projection of
  the magnetic field on the path. The electron density near the Sun
  can be measured by several methods, but it is quite variable.  Here
  we show that it is possible to measure the path integrated electron
  density and the Faraday rotation simultaneously at 6-10 $R_\odot$
  using millisecond pulsars as the linearly polarized radio
  source. 
  By analyzing the Faraday rotation measurements with and
  without the simultaneous electron density observations we show that
  these observations significantly improve the accuracy of the
  magnetic field estimates.
  
\end{abstract}


\section{Introduction}

The magnetic field in the solar corona and the inner solar wind is
very important because it controls the structure and dynamics of the
coronal plasma. Unfortunately it is very difficult to measure the
magnetic field in this region directly.  Optical observations can be
made in the photosphere, and these can be extrapolated outwards into
the corona \citep{swn69,an69}, but such extrapolations do not agree
well with estimates made by extrapolating inwards space-craft
measurements at Mercury \citep{bur01}.

It is possible to estimate the magnetic field in the quiet solar wind
by measuring the Faraday rotation of linearly polarized radio waves
propagating through the region of interest. Thus a good deal of work
has been done using the Faraday rotation technique
\citep[e.g.][]{lss+69,pbv+87,be90,ss94a,ss94b,ms99,ms00,jba+05,spa05,isw07}. The
Faraday rotation of the position angle of the linear polarization (in
cgs units) is given by
\begin{equation}
  \Delta\psi =  \lambda^2 \left [\frac{e^3}{2 \pi m_e^2 c^4} \int_{path} \!\!\!\!\!\! n_e(l)
  \mbox{\boldmath $B$} (l) \mbox{\boldmath $\cdot$} \mbox{d\boldmath $l$} \right] .
\end{equation}
The magnetic field of the inner corona is dominated by closed loops
over active regions and open field lines over coronal holes
\citep{gf98}. Closed loops seldom extend past 2-3 $R_\odot$ and the
magnetic field, as defined by bright striations in white light
observations, appears to be radial outside of this distance
\citep{gf95}. At larger distances a tangential component builds up due
to rotation of the Sun (Hundhausen, 1972), but in the regions of
interest for Faraday rotation observations the magnetic field is
predominantly radial.  This has an important consequence for Faraday
rotation observations.  If the corona were spherically symmetric the
sign of \( \mbox{\boldmath$B \cdot$}\mbox{d\boldmath$l$}\) would
reverse at the closest point of approach and the Faraday rotation
would approach zero. So one cannot use the simple model of a
spherically symmetric solar wind - Faraday rotation is a measure of
the deviation from spherical symmetry. A further problem with
interpreting Faraday rotation observations is that the electron
density must be known. Fortunately the electron density can be
measured by several methods and its average behavior is reasonably
well understood \citep[e.g.][]{bvp+94,ghm96}. However it is time
variable, and the path integration must be modeled carefully to obtain
reliable estimates of the magnetic field.

With a pulsar one can also measure the group delay of the pulse
passing through the solar wind. This is given by
\begin{equation}
\tau_g = \lambda^2 \left [ \frac{e^2}{2\pi m_ec^3} \int_{path} \!\!\!\!\!\! n_e (l) \mbox{d}l \right ].
\end{equation}
We will show that this constraint is very valuable because it allows
us to estimate the ratio (F) of the instantaneous electron density
$n_e(t)$ to the average $\langle n_e \rangle$. In our observations
this factor lay in the range $0.64 < F < 1.77$. Correction of the
magnetic field estimate by this factor is a significant improvement.
However the group delay due to the solar wind is small and cannot be
measured with sufficient accuracy for most pulsars. Fortunately the
class of pulsars with millisecond periods (MSPs) has very high
rotational stability and can be timed with a precision of the order of
100 ns \citep[e.g.][]{man08}. This is sufficient to measure the solar
wind contribution \citep{yhc+07a}.  Simultaneous measurements of
Faraday rotation and group delay can also be performed with suitably
equipped spacecraft and such measurements have been very useful, but
they not been possible since the Helios probes which worked up to
1985 \citep{hbv+82}.

\begin{figure}
\includegraphics[width=80mm,angle=0]{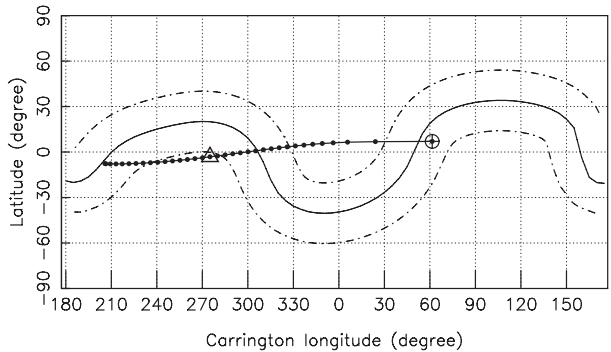}\label{fg:track}
\caption{The current sheet as determined by the WSO with the
  ``classic'' extrapolation to a source surface at 2.5 $R_\odot$ is
  shown as a heavy solid line. The location of the slow wind belt
  ($\pm20^\circ$ about the current sheet) is indicated by dashed
  lines. The path to PSR J1022+1001 on 2006 Aug 25 at 00:20 UT has
  been extrapolated back to the source surface along the streamlines
  assuming radial flow and a velocity of 400 km/s. The source-surface
  point corresponding to the point of closest approach to the Sun
  along the line of sight is indicated by a triangle and that
  corresponding to the Earth position by $\oplus$.  The furthest point
  from the Earth which is shown subtends an angle of 85$^\circ$ at the
  Sun.}
\end{figure}

Coronal modeling can be done most reliably near solar minimum when the
magnetic field structure is relatively simple and stable. At this
phase of the solar activity cycle it can be characterized by a
``magnetic pole'' which may be displaced from the rotational pole. The
magnetic field is assumed to be radial everywhere (and decreases with
distance like $R^{-2}$ to conserve flux) but is opposite in the
opposite hemispheres. Thus the magnetic equator is a current-sheet
where the field reverses. The current sheet appears above and below
the rotational equator as the Sun rotates. The solar wind velocity is
low and the electron density is high in a belt of about $\pm20^\circ$
around the current sheet. However, the magnitude of the magnetic field
is roughly the same in the fast and slow wind. Figure \ref{fg:track}
shows the location of the current sheet and the slow wind belt during
one of our observations.  Here the current sheet, as estimated using
the Wilcox Solar Observatory (WSO)\footnote{See
  http://wso.stanford.edu/} data, is shown as a heavy solid line and
the slow wind belt is bounded by dash-dotted lines. Points on the line
of sight from the Earth to the pulsar at the time of observation are
shown as a line of dots for which the spacing is 5$^\circ$ subtended
at the Sun. One can see that the line of sight crosses the current
sheet twice between the Earth and the closest point of approach and it
passes through two different low and high density regions. Thus there
is a current sheet at the magnetic equator where the field reverses.

However the location of the current sheet is subject to some error
because a current-free potential-field model was used to extrapolate the photospheric
observations of the WSO into the corona. Also the width of the slow wind
belt is not well-defined and the density may vary with time.
In this work we are concerned with the effect of these modeling errors on the
final magnetic field estimate, and how these can be reduced using the
additional constraint provided by a simultaneous group delay measurement.
This was suggested by \citet{ojs07}, but they were
unable to measure the group delay with sufficient accuracy in their
observations. We do not have enough measurements to make a
strong statement about the magnetic field of the solar wind, other than
to show consistency with other measurements, but we have sufficient data
to show the modeling errors one can expect and to compare them with
statistical errors. In the following sections we will discuss: the observations
and the primary analysis; the coronal model and the fitting process;
the magnetic fields estimates and their sensitivity to model errors;
and the prospects for future such observations.

\section{Observations and Primary Analysis}

The observations were part of the Parkes Pulsar Timing Array (PPTA)
project which commenced in 2004 and has made observations of 20
MSPs at intervals of 2-3 weeks since then
\citep{man08}. Here we restrict our analysis to the pulsar which
provides the best observations for our purpose, PSR J1022$+$1001 which
has ecliptic latitude $b_e = -0.06^\circ$, but two other MSPs
regularly observed with the PPTA, J1730$-$2304 ($b_e =
  0.19^\circ$) and J1824$-$2452 ($b_e = -1.55^\circ$) are potentially useful. The
  observations analysed are those when the line of sight to PSR
  J1022$+$1001 happened to be close to the Sun - they were not
  deliberately scheduled to be near the Sun. As a result there are few
  observations useful for this work.  All observations use the center
  beam of the Parkes 20-cm Multibeam receiver \citep{swb+96} at
  frequencies close to 1400\,MHz. All of the data were recorded in
  full-polarization mode. The back-end signal processing systems are
  Parkes digital filter bank systems (PDFB1 to PDFB4). Each
  observation has a total bandwidth of 256\,MHz with 1024
  channels. The duration of each observation is 64 minutes and the
  data were averaged with 1-min sub-integrations. A 2-min pulsed
  calibration signal was recorded before each pulsar observation. The
  flux density scale was set using observations of Hydra A. The
  cross-coupling between feed probes was measured using observations
  of PSR J0437$-$4715 covering a wide range of hour angles.

The primary analysis was done using \textsc{PSRCHIVE}
software \citep{hvm04}.  Initially, 5\% of the band edges and
radio frequency interference (RFI) were automatically removed.
Then the data were flux- and polarization-calibrated and the cross-coupling
between the feed polarizations was corrected using the \textsc{pac/pcm} algorithm in
\textsc{PSRCHIVE}. In pulsar analysis the group delay is quantified by the
``dispersion measure'' DM which is the path integrated electron density in pc cm$^{-3}$.
The Faraday rotation is quantified by the ``rotation measure'' RM, which is the
term in square brackets in equation 1 in units of rad m$^{-2}$.
We will use these measures in our further discussion.

The method of obtaining RM is similar to that discussed
in \citet{ymv+11}. We approached the final result with a series of three successive
approximations. We first adjusted the RM to optimize the linearly polarized
intensity integrated over the bandwidth. Then we separated each
observation into upper and lower halves of the total band. An improved
value of RM was determined using the weighted mean of position
angle difference between the two halves of the band for each pulse phase bin. The
RM also includes a time variable ionospheric component. We
corrected it using an ionospheric model which is encoded in 2007 International
Reference Ionophere (IRI) model\footnote{See http://iri.gsfc.nasa.gov/}.
It was relatively small, ranging from $-$0.45 to $-$2.25\,rad\,m$^{-2}$, but
not negligible.

To obtain the final approximation to RM we used the corrected position
angle vs pulse phase measured far from the Sun by \citet{ymv+11} as a
reference template.  We then found a weighted average of the
differences of each observation near the Sun to this reference
template.  The normalized chi-squared of these weighted averages was
typically about 1.3, indicating that the error estimates used in the
weighted fit were reasonably accurate and the position-angle template
was a good model when the pulsar was near the Sun.

To estimate the DM contribution from the solar wind, we used a simpler procedure.
The DM variations due to the interstellar medium for PSR J1022$+$1001 are smaller than
the measurement error \citep{yhc+07b}. So we simply
estimated pulse arrival times in the normal way, by fitting a detailed
timing model to the entire set of observations, excluding those near the Sun.
We then obtained DM$_\odot$ by comparing the observed timing residual near
the Sun with this distant average.
There is a contribution from the ionosphere but, unlike the RM contribution, the
ionospheric DM contribution is negligible compared with the measurement error.

\begin{figure}
\centerline{\includegraphics[width=80mm,angle=0]{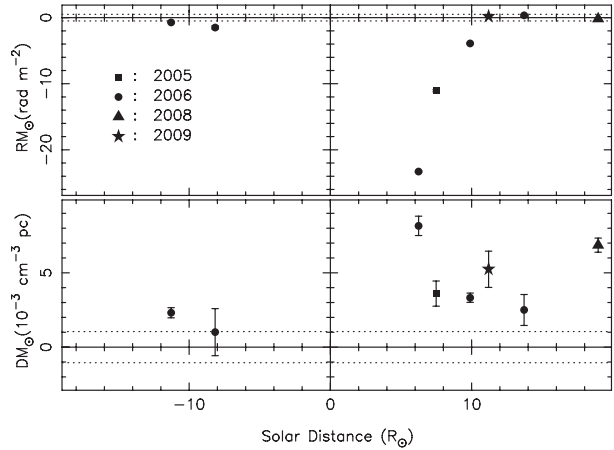}}
\caption{Observations of PSR J1022+1001 near the Sun. The abscissa is the distance from the Sun. The E limb
data is plotted with a negative distance, and the W limb data with a positive distance.
The observations for different years are marked by different symbols.
The error bars on the RM are smaller than the symbol size.}\label{fg:rmdm}
\end{figure}

The final RM and DM measurements used are summarized in Table \ref{tb:obs} and
plotted in Figure \ref{fg:rmdm} vs distance from the Sun.

\begin{table*}
\caption{Observations of the PSR J1022$+$1001 near the Sun.
The ecliptic coordinates are $b_e = -0.06^\circ$ and $l_e = 153.9^\circ$.}\label{tb:obs}
\begin{tabular}{cccccccccc}
\hline
 Date & UT & $R_{0}$      &   $I$    & $L/I$ & RM$_\odot$      &  DM$_\odot$ \\
         &      & (R$_{\odot}$) & (mJy)        & (\%)  & (rad\,m$^{-2}$) &  (10$^{-3}$cm$^{-3}$pc) \\
\hline
 2005/08/29 & 02:15 & 7.5   & 3.0  $\pm$0.3  & 62 & $-11.6\pm0.15$ & $3.6\pm0.8$ \\
 2006/08/24 & 03:40 & $-$11.3 & 8.0  $\pm$0.4  & 52 & $-0.72\pm0.06$ & $1.2\pm0.3$  \\
 2006/08/25 & 00:20 & $-$8.2  & 4.6  $\pm$1.0 & 48 & $-1.49\pm0.24$ & $0.3\pm2.0$\\
  2006/08/28 & 23:49 & 6.2   & 13.3 $\pm$1.3  & 58 &$-23.3\pm0.09$ & $8.2\pm0.6$ \\
 2006/08/30 & 00:02 & 9.9   & 12.4 $\pm$0.7  & 52 & $-3.9\pm0.05$ & $3.6\pm0.3$ \\
 2006/08/31 & 01:19 & 13.7  & 2.3  $\pm$0.4  & 54 & $0.34\pm0.16$ & $2.0\pm1.3$ \\
  2008/08/31 & 23:41 & 19.0  & 1.1  $\pm$0.2  & 48 & $-0.18\pm0.17$ & $5.6\pm0.9$ \\
   2009/08/30 & 02:01 & 11.2  & 3.2  $\pm$0.2  & 55 & $0.17\pm0.08$ & $7.7\pm0.4$ \\
\hline
\end{tabular}
\end{table*}

\section{Solar wind model fitting}

We assume that the electron density is bimodal, as noted earlier.  The
average density in the two modes has been estimated
\citep{gf95,gf98,ma81,all47}; we use the same approximations as were
used by \citet{yhc+07a}.  The density $n_e$ in the fast wind is
\begin{equation}
n_e (R) = (1.155 R^{-2} + 32.2 R^{-4.39} + 3254 R^{-16.25} ) 10^{11}\;{\rm m}^{-3}
\end{equation}
and $n_e$ in the slow wind is
\begin{eqnarray}
n_e (R) =& (4.1 R^{-2} + 23.53 R^{-2.7}) 10^{11} +  \nonumber \\
& ( 1.5 R^{-6} + 2.99 R^{-16} )10^{14}\;{\rm m}^{-3}.
\end{eqnarray}
Here $R$ is the distance from the center of the Sun in $R_\odot$.
To match the DM observations we scale the electron density by a factor F, and to
minimize the number of free parameters we use the same factor in both the fast and slow wind.

We assume that the slow wind belt is centered on the current sheet as
determined by the WSO photospheric observations extrapolated into the
corona \citep{mbf+00}. However the width of the slow wind belt is
variable and not precisely determined in any case \citep{ms00}.  Accordingly we
have tested models with widths of $\pm10^\circ$, $\pm20^\circ$ and
$\pm30^\circ$. 

The extrapolation is done with three similar, but
subtly different, techniques referred to as: ``classic''; ``radial
250''; and ``radial 325'' in the WSO data base\footnote{see
  http://wso.stanford.edu/synsourcel.html}. The three models find a
potential solution which is current free and radial at some source
surface.  The radial models both constrain the photospheric field to
be radial and differ only in the source surface which is at 2.50 or
3.25 $R_\odot$. The classic model does not constrain the photospheric
field to be radial but requires a somewhat ad hoc polar field
correction. It assumes a source surface at 2.5 $R_\odot$.  We
assume that the magnetic field is the same in the fast and slow wind,
that it is radial, and that it varies quadratically with solar
distance.  We take the polarity of the magnetic field from the
corresponding WSO model.

\begin{figure}
\centerline{\includegraphics[width=82mm,angle=0]{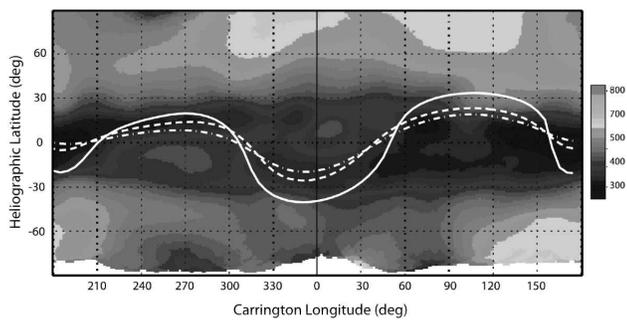}}

\caption{Solar wind velocity from STELAB, Nagoya U. This figure has
  been rotated in longitude to match Figure \ref{fg:track}.  The
  current sheet locations from the three WSO models have been
  overlaid. The ``classic'' model is a solid line, the ``radial 250''
  model is dashed, and the ``radial 325'' model is
  dot-dashed.}\label{fg:ste}
\end{figure}

To provide an idea of the range of the extrapolation models and a
comparison with other solar wind observations, we plot the current
sheets for the three models over the solar wind velocity measurements
for 2006 from the
STELAB\footnote{See http://www.stelab.nagoya-u.ac.jp/omosaic/crle.html} in
Figure \ref{fg:ste}. Here one can see that the three extrapolation models vary
significantly in the maximum latitude extent of the current sheet, and
that the width of the slow velocity belt is not clearly defined.

\begin{figure}
\centerline{\hspace{4cm}\includegraphics[width=110mm,angle=0]{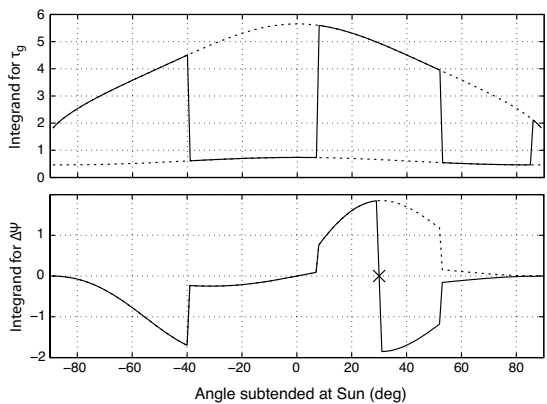}}
\caption{Integrands for group delay (top panel) and Faraday rotation
  (bottom panel) for the path shown in Figure \ref{fg:track}.  The
  integrands for fast and slow wind alone are shown dashed in the top
  panel.  The integrand for Faraday rotation if the phase reversal at
  the current sheet were not present is shown dashed in the lower
  panel. The $\times$ marks the point near Carrington longitude
  $310\degr$ where the line of sight crosses the current
  sheet (cf. Figure~\ref{fg:track}).}\label{fg:ker}
\end{figure}

It is helpful to examine the integrand of the path integrals to see
where the maximum contributions arise.  For numerical purposes we
transform the path integral into an integral over the angle subtended
at the Sun. This makes the range finite and compresses the integrand
in a useful way.  The integrands for group delay and Faraday rotation
for the path shown in Figure \ref{fg:track} are shown in Figure
\ref{fg:ker}.  One can see that the contributions from the slow dense
wind are dominant in both integrands and the location of the current
sheet is very important in obtaining the Faraday rotation.

The fitting procedure consists of three steps. First, we examine the
LASCO C3 movies\footnote{See
  http://soho.esac.esa.int/data/archive/index\_ssa.html} on the three
days surrounding the observations and confirm that there was no obvious
transient during the observing period. This was true of all the
observations in Table \ref{tb:obs}. Second, we calculate the DM from one of the 9
models (3 extrapolation methods and 3 widths of the slow wind belt)
and find the factor F necessary to match the observed DM in each
case. Third, we use that factor F and an arbitrary $|B|$ to calculate
the RM for that model, and scale $|B|$ until the model matches the
observed RM. The DM and RM steps are repeated for all 9 models.

There were three observations that we could not model completely in
this way.  In one case, 2006 Aug 25, the DM measurement is not
adequate to determine the factor F, so we set F=1.  In the second
case, 2009 Aug 30, all three models showed that the path was entirely
in the slow wind. The electron density was 20\% above its average (F =
1.2), but the RM was very small - within 2 $\sigma$ of zero. This
would be expected if the path did not cross the current sheet. Two of
the three extrapolation models indicated that the path would cross the
current sheet and give a significant RM, but the third, the ``radial
250'' model did not indicate a current sheet crossing so the predicted
RM = 0. In this case we have a consistent model but no estimate of the
magnetic field.  In the third case, 2008 Aug 31, all three models show
that the path is almost entirely in the slow wind. The electron
density was high (F = $3.4\pm0.4$) and the RM was very small. The
``radial 325'' model predicted the wrong sign of RM indicating a
significant error in the location of the current sheet. The
``classic'' model predicted $|B| = 0.2 \pm 0.2$ mG and the ``radial
250'' model $|B| = 59 \pm 56$ mG. We doubt that $|B| < 0.4$ mG at 11.2
R$_\odot$ and suggest that the ``radial 250'' model is more consistent
with the observation.

The results for the nine different models are shown in Table
\ref{tb:sens}.  One can see that the factor F varies from 0.5 to 3.0,
which confirms that having simultaneous DM observations provides a
strong constraint on the solar wind model.  We have taken the mean of
the 9 estimates of $|B|$ for each observation as the best estimate of
$|B|$ for a given observation. The rms of these 9 estimates is a
measure of the model sensitivity of that particular estimate. We also
have an estimate of the statistical error derived from propagation of
the errors on the primary measurements of DM and RM. These statistical
errors are dominated by the error in measuring DM. Finally we find the
$|B|$ estimate that would have been made in the absence of DM
observations simply by setting F = 1 for each model and recalculating
$|B|$. This summary is shown in Table \ref{tb:sta}. Here one can see
that the model errors (merr) exceed the statistical (sterr) errors in every
case, and the average effect of correcting for the observed DM 
substantially exceeds the model errors.

\begin{table}
\caption{Results of model sensitivity analysis using 3 WSO models and
  3 widths for slow wind. Here $|B|$ in mG is estimated at the point where 
  the line of sight is closest to the Sun.}\label{tb:sens}
\begin{tabular}{lccccccc}
\hline
Observation &  mod & \multicolumn{2}{|c|}{$\pm10^{\circ}$} & \multicolumn{2}{c}{$\pm20^{\circ}$} & \multicolumn{2}{c}{$\pm30^{\circ}$} \\
      &  & F & $|B|$ & F & $|B|$ & F & $|B|$ \\
\hline
2005/08/29  &  clas& 0.9 & 48.9 & 0.6 & 33.1 & 0.5 & 26.6 \\
02:16 UT        & 250  &0.9  & 29.6 & 0.6 & 23.2 & 0.5 & 20.0 \\
7.5 $R_\odot$        & 325  & 0.8 & 23.7 & 0.5 & 19.3 & 0.5 & 19.0 \\
\hline
2006/08/28    & clas & 3.0 & 24.2 & 2.1 & 21.9 & 1.3  & 19.8  \\
23:49 UT          & 250 & 2.7 & 21.8 & 1.4 & 17.2 & 0.9 & 14.8  \\
6.2 $R_\odot$      & 325 & 2.4& 20.8 & 1.3 & 16.4 & 0.8 & 15.0 \\
\hline
2006/08/30  & clas & 2.3 & 12.0 & 1.4 & 12.7 & 1.0 & 10.4 \\
00:02 UT        & 250 & 1.7 & 10.2 & 1.0 & 8.2 & 0.7 & 6.5 \\
9.9 $R_\odot$ & 325 & 1.8 & 8.4  & 0.9 & 7.4  & 0.6 & 6.7  \\
\hline
2006/08/25 & clas & 1.0 & 14.1 & 1.0 & 3.5 & 1.0 & 2.7 \\
20:56 UT        & 250 & 1.0 & 28.5 & 1.0& 3.4 & 1.0 & 3.4 \\
$-8.2 R_\odot$ & 325 & 1.0 & 10.6 & 1.0 & 2.3    & 1.0 & 1.2 \\
\hline
\end{tabular}
\end{table}

\begin{table*}
\caption{Final estimates of $|B|$ at the point where 
  the line of sight is closest to the Sun.  $|B|$  is given both corrected
  by measured DM and uncorrected. The measured DM exceeds that
  expected from the electron density model by the factor F.  }\label{tb:sta}
\begin{tabular}{cc|cccc|ccc}
\hline
Observation & $R_{0}$ & \multicolumn{4}{c}{$|B|$ (mG)}
&\multicolumn{3}{c}{F}\\ \cline{3-6}
	& ($R_\odot$) & DMcor & uncor &merr &sterr & mean & merr & sterr\\
\hline
2005/08/29  & 7.5 & 27.0 &42.3 & $\pm9.5$ & $\pm6.7$ & 0.64 &$\pm0.17$ &$\pm0.14$ \\

2006/08/28 & 6.2 & 19.1 &12.3 & $\pm3.4$ & $\pm1.4$ & 1.77 &$\pm0.80$ &$\pm0.13$  \\

2006/08/30 & 9.9 & 9.2 &8.0 & $\pm2.3$ & $\pm1.1$ & 1.27&$\pm0.57$ &$\pm0.11$  \\

2006/08/25 & $-8.2$ &    &7.7& $\pm8.9$ & $\pm1.5$ & 1.0 & & \\
\hline
\end{tabular}
\end{table*}

\section{Discussion of results}

We have identified four primary contributions to the error on magnetic field estimates
made using the Faraday rotation technique. In order of importance these are:
failure to correct for DM variations ($\approx$50\%); inaccuracy of
estimation of the current sheet location ($\approx$25\%); error in the DM measurement
($\approx$15\%); and error in the Faraday rotation estimate ($\approx$1\%).
Thus use of simultaneous DM observations can roughly double the precision of the magnetic
field estimates. The value of the PPTA for such measurements is clear. It
provides a good timing model for the DM far from the Sun
and a good template for the position-angle variations far from the Sun.

\begin{figure}
\centerline{\includegraphics[width=90mm,angle=0]{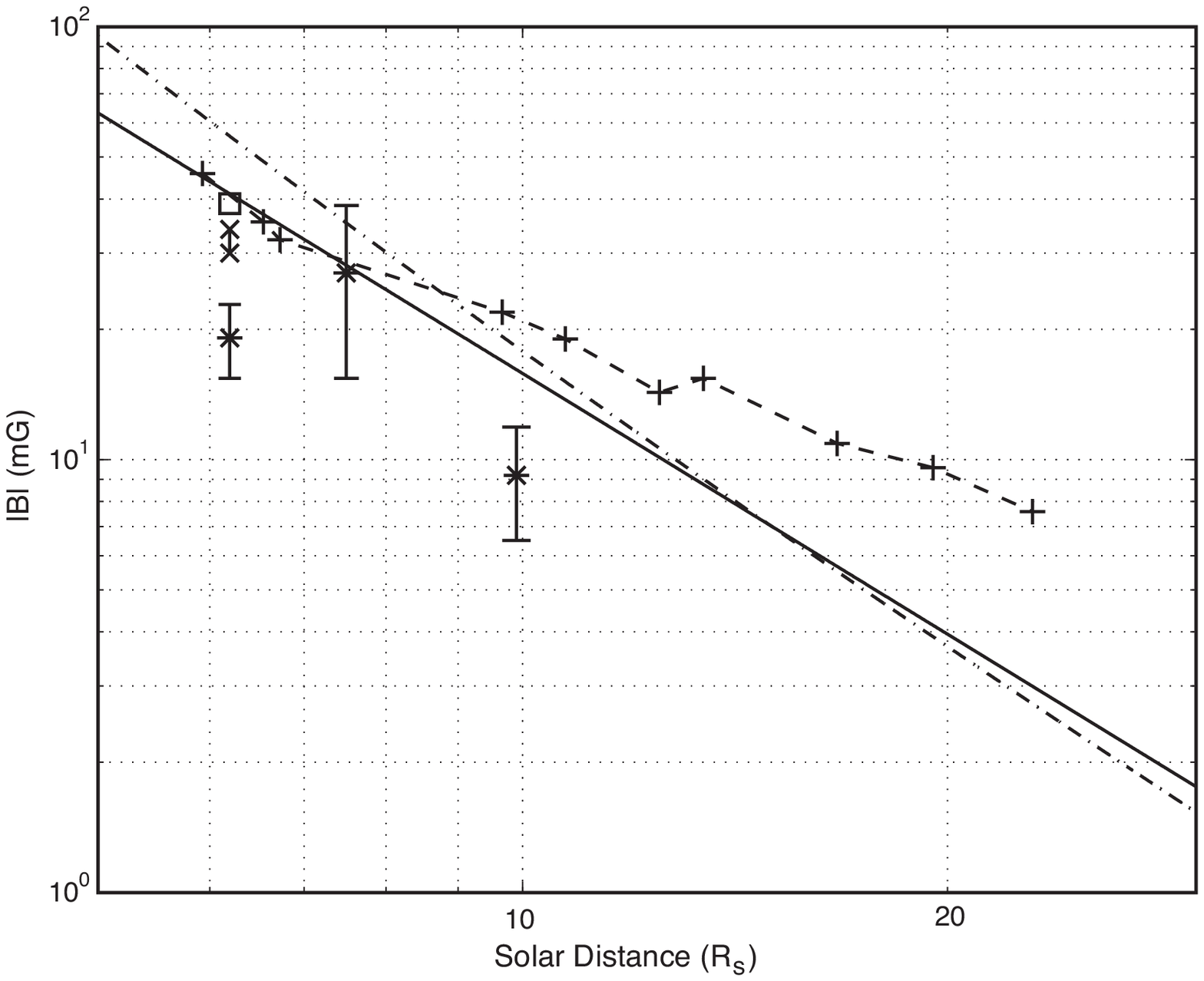}}
\caption{Magnetic field observations near the Sun. The measurements in
  Table \ref{tb:sta} are plotted as * symbols with error bars.  The
  extrapolated Helios measurements are a solid line \citep{bur01}, an
  average of the \citet{pbv+87} observations ($|B| = 1000 (6 R^{-3} +
  1.18 R^{-2})$ mG) are a dot-dashed line and the observations of
  \citet{gy11} are + symbols connected by a dashed line.  The
  observation of \citet{spa05} is an open box and the range given by
  \citet{isw07} is a pair of connected x symbols.  }\label{fg:mag}
\end{figure}

We have plotted our estimates of $|B|$ along with those of other
authors in Figure \ref{fg:mag}.  The estimates of \citet{ojs07} and
\citet{scs09} are not plotted because in both cases the authors only
provide a measurement of the average line-of-sight component of the
magnetic field and do not attempt to estimate $|B|$. Their estimates
of the average line-of-sight components are consistent with other
observations but do not put a useful constraint on $|B|$.
Our estimates of $|B|$ are somewhat lower than the others, but are
consistent with them, within the error bars. All the observations
are broadly consistent with the Helios space craft measurements and 
an extrapolation like $R^{-2}$. Thus our measurements, although few
in number, confirm the very important result that the potential field
models of the corona seriously underestimate the magnitude of the 
magnetic field for $R > 5 R_\odot$.
 
We did not observe any transients near the Sun and would not have been
able to analyze them with the technique we used.  However it might be
possible to do so if sufficient additional information, such as might
be provided by STEREO,\footnote{See http://stereo.gsfc.nasa.gov} were
available to constrain the geometry of the magnetic field.
We note that Faraday rotation is capable of 1\% precision by itself, and
one might be able to measure rapid variations of magnetic field during 
transients with this accuracy.

We believe that more extensive observations of PSRs J1022$+$1001,
J1730$-$2304, and J1824$-$2452 near the Sun, perhaps involving more
telescopes to provide continuous coverage during the closest approach,
would be valuable and should be undertaken at the next
opportunity. There are five additional MSPs with ecliptic latitudes
$\le 2^\circ$ that may be suitable for precision timing: J0030$+$0451;
J1614$-$2230; J1721$-$2457; J1802$-$2124; and J1811$-$24.  Preliminary
observations would be required to confirm that these are ``good
timers'' before undertaking coronal measurements. The cluster of
pulsars with a right ascension close to 18:00 would facilitate an
intensive observing program in the last two weeks of December each
year, but there are insufficient MSPs to monitor the corona on a
regular basis.

\section*{Acknowledgments}

XPY is supported by the National Natural Science Foundation of China
(10803004), the Natural Science Foundation Project of ChongQing (CQ
CSTC 2008BB0265) and the Fundamental Research Funds for the Central
Universities (XDJK2012C043). GH acknowledges support from the Chinese
Academy of Sciences CAS KJCX2-YW-T09 and NSFC 11050110423. The data
presented in this paper were obtained as part of the Parkes Pulsar
Timing Array project. We thank the collaborators on this project. The
Parkes radio telescope is part of the Australia Telescope which is
founded by the Commonwealth of Australia for operation as a National
Facility managed by CSIRO.


\begin{thebibliography}{}

\bibitem[\protect\citeauthoryear{{Allen}}{{Allen}}{1947}]{all47}
{Allen} C.~W.,  1947, MNRAS, 107, 426

\bibitem[\protect\citeauthoryear{{Altschuler} \& {Newkirk}}{{Altschuler} \&
  {Newkirk}}{1969}]{an69}
{Altschuler} M.~D.,  {Newkirk} G.,  1969, Solar Phys., 9, 131

\bibitem[\protect\citeauthoryear{{Bird} \& {Edenhofer}}{{Bird} \&
  {Edenhofer}}{1990}]{be90}
{Bird} M.~K.,  {Edenhofer} P.,  1990, {Remote Sensing Observations of the Solar
  Corona}.
p.~13

\bibitem[\protect\citeauthoryear{{Bird}, {Volland}, {Paetzold}, {Edenhofer},
  {Asmar} \& {Brenkle}}{{Bird} et~al.}{1994}]{bvp+94}
{Bird} M.~K.,  {Volland} H.,  {Paetzold} M.,  {Edenhofer} P.,  {Asmar} S.~W.,
   {Brenkle} J.~P.,  1994, ApJ, 426, 373

\bibitem[\protect\citeauthoryear{{Burlaga}}{{Burlaga}}{2001}]{bur01}
{Burlaga} L.~F.,  2001, Planetary and Space Science, 49, 1619

\bibitem[\protect\citeauthoryear{{Gopalswamy} \& {Yashiro}}{{Gopalswamy} \&
  {Yashiro}}{2011}]{gy11}
{Gopalswamy} N.,  {Yashiro} S.,  2011, ApJ, 736, L17+

\bibitem[\protect\citeauthoryear{{Guhathakurta} \& {Fisher}}{{Guhathakurta} \&
  {Fisher}}{1998}]{gf98}
{Guhathakurta} M.,  {Fisher} R.,  1998, ApJ, 499, L215+

\bibitem[\protect\citeauthoryear{{Guhathakurta} \& {Fisher}}{{Guhathakurta} \&
  {Fisher}}{1995}]{gf95}
{Guhathakurta} M.,  {Fisher} R.~R.,  1995, Geophys. Res. Lett., 22, 1841

\bibitem[\protect\citeauthoryear{{Guhathakurta}, {Holzer} \&
  {MacQueen}}{{Guhathakurta} et~al.}{1996}]{ghm96}
{Guhathakurta} M.,  {Holzer} T.~E.,    {MacQueen} R.~M.,  1996, ApJ, 458, 817

\bibitem[\protect\citeauthoryear{{Hollweg}, {Bird}, {Volland}, {Edenhofer}, {Stelzried} \&
  {Seidel}}{{Hollweg} et~al.}{1982}]{hbv+82}
{Hollweg} J.~V., {Bird} M.~K., {Volland} H., {Edenhofer} P.,
	{Stelzried} C.~T., {Seidel}, B.~L.,  1982, J. Geophys. Res., 87, 1

\bibitem[\protect\citeauthoryear{{Hotan}, {van Straten} \&
  {Manchester}}{{Hotan} et~al.}{2004}]{hvm04}
{Hotan} A.~W.,  {van Straten} W.,    {Manchester} R.~N.,  2004, PASA, 21, 302

\bibitem[\protect\citeauthoryear{{Ingleby}, {Spangler} \& {Whiting}}{{Ingleby}
  et~al.}{2007}]{isw07}
{Ingleby} L.~D.,  {Spangler} S.~R.,    {Whiting} C.~A.,  2007, ApJ, 668, 520

\bibitem[\protect\citeauthoryear{{Jensen}, {Bird}, {Asmar}, {Iess}, {Anderson}
  \& {Russell}}{{Jensen} et~al.}{2005}]{jba+05}
{Jensen} E.~A.,  {Bird} M.~K.,  {Asmar} S.~W.,  {Iess} L.,  {Anderson} J.~D.,
   {Russell} C.~T.,  2005, Adv. Space Res., 36, 1587

\bibitem[\protect\citeauthoryear{{Levy}, {Sato}, {Seidel}, {Stelzried},
  {Ohlson} \& {Rusch}}{{Levy} et~al.}{1969}]{lss+69}
{Levy} G.~S.,  {Sato} T.,  {Seidel} B.~L.,  {Stelzried} C.~T.,  {Ohlson} J.~E.,
     {Rusch} W.~V.~T.,  1969, Science, 166, 596

\bibitem[\protect\citeauthoryear{{Manchester}}{{Manchester}}{2008}]{man08}
{Manchester} R.~N.,  2008, in {C.~Bassa, Z.~Wang, A.~Cumming, \& V.~M.~Kaspi}
  ed., 40 Years of Pulsars: Millisecond Pulsars, Magnetars and More Vol.~983 of
  American Institute of Physics Conference Series, {The Parkes Pulsar Timing
  Array Project}.
pp 584--592

\bibitem[\protect\citeauthoryear{{Mancuso} \& {Spangler}}{{Mancuso} \&
  {Spangler}}{1999}]{ms99}
{Mancuso} S.,  {Spangler} S.~R.,  1999, ApJ, 525, 195

\bibitem[\protect\citeauthoryear{{Mancuso} \& {Spangler}}{{Mancuso} \&
  {Spangler}}{2000}]{ms00}
{Mancuso} S.,  {Spangler} S.~R.,  2000, ApJ, 539, 480

\bibitem[\protect\citeauthoryear{{McComas}, {Barraclough}, {Funsten},
  {Gosling}, {Santiago-Mu{\~n}oz}, {Skoug}, {Goldstein}, {Neugebauer}, {Riley}
  \& {Balogh}}{{McComas} et~al.}{2000}]{mbf+00}
{McComas} D.~J.,  {Barraclough} B.~L.,  {Funsten} H.~O.,  {Gosling} J.~T.,
  {Santiago-Mu{\~n}oz} E.,  {Skoug} R.~M.,  {Goldstein} B.~E.,  {Neugebauer}
  M.,  {Riley} P.,    {Balogh} A.,  2000, J. Geophys. Res., 105, 10419

\bibitem[\protect\citeauthoryear{{Muhleman} \& {Anderson}}{{Muhleman} \&
  {Anderson}}{1981}]{ma81}
{Muhleman} D.~O.,  {Anderson} J.~D.,  1981, ApJ, 247, 1093

\bibitem[\protect\citeauthoryear{{Ord}, {Johnston} \& {Sarkissian}}{{Ord}
  et~al.}{2007}]{ojs07}
{Ord} S.~M.,  {Johnston} S.,    {Sarkissian} J.,  2007, Solar Phys., 245, 109

\bibitem[\protect\citeauthoryear{{Patzold}, {Bird}, {Volland}, {Levy}, {Seidel}
  \& {Stelzried}}{{Patzold} et~al.}{1987}]{pbv+87}
{Patzold} M.,  {Bird} M.~K.,  {Volland} H.,  {Levy} G.~S.,  {Seidel} B.~L.,
  {Stelzried} C.~T.,  1987, Solar Phys., 109, 91

\bibitem[\protect\citeauthoryear{{Sakurai} \& {Spangler}}{{Sakurai} \&
  {Spangler}}{1994a}]{ss94a}
{Sakurai} T.,  {Spangler} S.~R.,  1994a, ApJ, 434, 773

\bibitem[\protect\citeauthoryear{{Sakurai} \& {Spangler}}{{Sakurai} \&
  {Spangler}}{1994b}]{ss94b}
{Sakurai} T.,  {Spangler} S.~R.,  1994b, Rad. Sci., 29, 635

\bibitem[\protect\citeauthoryear{{Schatten}, {Wilcox} \& {Ness}}{{Schatten}
  et~al.}{1969}]{swn69}
{Schatten} K.~H.,  {Wilcox} J.~M.,    {Ness} N.~F.,  1969, Solar Phys., 6, 442

\bibitem[\protect\citeauthoryear{{Smirnova}, {Chashei} \& {Shishov}}{{Smirnova}
  et~al.}{2009}]{scs09}
{Smirnova} T.~V.,  {Chashei} I.~V.,    {Shishov} V.~I.,  2009, Astronomy
  Reports, 53, 252

\bibitem[\protect\citeauthoryear{{Spangler}}{{Spangler}}{2005}]{spa05}
{Spangler} S.~R.,  2005, Space Sci. Rev., 121, 189

\bibitem[\protect\citeauthoryear{Staveley-Smith, Wilson, Bird, Disney, Ekers,
  Freeman, Haynes, Sinclair, Vaile, Webster \& Wright}{Staveley-Smith
  et~al.}{1996}]{swb+96}
Staveley-Smith L.,  Wilson W.~E.,  Bird T.~S.,  Disney M.~J.,  Ekers R.~D.,
  Freeman K.~C.,  Haynes R.~F.,  Sinclair M.~W.,  Vaile R.~A.,  Webster R.~L.,
    Wright A.~E.,  1996, PASA, 13, 243

\bibitem[\protect\citeauthoryear{{Yan}, {Manchester}, {van Straten},
  {Reynolds}, {Hobbs}, {Wang}, {Bailes}, {Bhat}, {Burke-Spolaor}, {Champion},
  {Coles}, {Hotan}, {Khoo}, {Oslowski}, {Sarkissian}, {Verbiest} \&
  {Yardley}}{{Yan} et~al.}{2011}]{ymv+11}
{Yan} W.~M.,  {Manchester} R.~N.,  {van Straten} W.,  {Reynolds} J.~E.,
  {Hobbs} G.,  {Wang} N.,  {Bailes} M.,  {Bhat} N.~D.~R.,  {Burke-Spolaor} S.,
  {Champion} D.~J.,  {Coles} W.~A.,  {Hotan} A.~W.,  {Khoo} J.,  {Oslowski} S.,
   {Sarkissian} J.~M.,  {Verbiest} J.~P.~W.,    {Yardley} D.~R.~B.,  2011,
  MNRAS, 414, 2087

\bibitem[\protect\citeauthoryear{{You}, {Hobbs}, {Coles}, {Manchester} \&
  {Han}}{{You} et~al.}{2007a}]{yhc+07a}
{You} X.~P.,  {Hobbs} G.~B.,  {Coles} W.~A.,  {Manchester} R.~N.,    {Han}
  J.~L.,  2007a, ApJ, 671, 907

\bibitem[\protect\citeauthoryear{{You}, {Hobbs}, {Coles}, {Manchester},
  {Edwards}, {Bailes}, {Sarkissian}, {Verbiest}, {van Straten}, {Hotan}, {Ord},
  {Jenet}, {Bhat} \& {Teoh}}{{You} et~al.}{2007b}]{yhc+07b}
{You} X.~P.,  {Hobbs} G.,  {Coles} W.~A.,  {Manchester} R.~N.,  {Edwards} R.,
  {Bailes} M.,  {Sarkissian} J.,  {Verbiest} J.~P.~W.,  {van Straten} W.,
  {Hotan} A.,  {Ord} S.,  {Jenet} F.,  {Bhat} N.~D.~R.,    {Teoh} A.,  2007b,
  MNRAS, 378, 493

\end{thebibliography}
\end{document}